\documentclass[prl,tightenlines,twocolumn]{revtex4}

\usepackage{graphicx,epsfig}
\usepackage{bm}
\usepackage{latexsym,amssymb,amsmath,float}

\def\nn{\nonumber}
\def\beq{\begin{eqnarray}}
\def\eeq{\end{eqnarray}}

\def\no{\urcorner}
\def\cc{\Lambda}
\def\life{\textrm{life}}
\def\us{\textrm{us}}
\def\a{\alpha}
\def\b{\beta}
\def\g{\gamma}
\def\e{\epsilon}

\def\P{{\cal P}}

\begin{document}

\title{Anthropics and Myopics: Conditional Probabilities and the Cosmological Constant}

\author{Irit Maor$^1$, Lawrence Krauss$^{1,2}$ and Glenn Starkman$^{1,2}$}
\affiliation{
 $^1$ CERCA, Department of Physics, Case Western Reserve University,
   10900 Euclid Avenue, Cleveland, OH 44106-7079, USA.\\
 $^2$ also Department of Astronomy, Case Western Reserve University}

\begin{abstract}
We re-examine claims that anthropic arguments provide an explanation for the observed smallness of the cosmological constant, and argue that correlations between the cosmological constant value and the existence of life can be demonstrated only under restrictive assumptions. Causal effects are more subtle to uncover.
\end{abstract}

%\date{\today}
\maketitle

%%%%%%%%%%%%%%%%%%%%%%%%%%%%%%%%%%%%%%%%%%%%%%%%%%%%%%%%%%%%%%%%%%%%%%%%%

The argument that if the cosmological constant (CC) were much larger than the value inferred from observations then galaxies wouldn't have formed and life would not exist \cite{wein,etc} is by now well known.
The lack of any compelling explanation on the basis of fundamental physics for why the vacuum energy today should be small has led this anthropic `explanation' to steadily gain favor, provoking significant work on a so-called ``landscape" analysis of possible vacua in string theory models, for example.
The anthropic principle is however based fundamentally on ignorance rather than knowledge; thus any probabilistic inferences one draws using it should be carefully interpreted.
Not only do we have no fundamental theory that might provide an underlying probability distribution for universes with different CCs, we know very little about the range of fundamental parameters that might allow the evolution of intelligent life in a universe.
In this paper we illustrate a point made very clearly by Weinberg in his original discussion on the subject \cite{wein}, namely that  a correlation between a universe with life as we know it, and the existence of a small CC need not be understood to imply causation, or as he put it, our existence as an experimental datum has no real explanatory power. We further argue that any inferred correlation between the CC and life strongly depends on how life is defined. Within the framework of the multiple vacua suggested by the landscape of string theory, predictive power about the parameters of our universe is often claimed to be obtainable through probabilistic anthropic reasoning about the {\it observability} of these parameters. However, arguments of these sort (i.e. see \cite{bousso}) strongly rely on the assertion that we must be typical observers \cite{page}, an assertion without sound fundamental scientific basis at the current time \cite{h+s}.

\noindent{{\bf{Life states}}}: \\
A universe can be categorized by whether it sustains intelligent life. For the purpose of this paper we define ``intelligent life" as a life form capable of measuring the CC and pondering its value. In other words, we define ``life" as an organism capable of producing scientists. If life exists, it can be further divided into those life forms which are sufficiently similar to us that we can estimate the necessary conditions for existence {\it etc}, and those about which we know nothing. Thus a universe can be in one of 3 states: {\bf 1)}
$\no\life$: There isn't any scientist-producing life whatsoever in the universe. We call this state ``not life".
{\bf 2)}
$\no\us$: There is some life, but one which isn't recognizable to us, ``not us". We don't know what it needs in order to exist and survive. For example a life form may
thrive  in a universe where the particle symmetry is {\it not}
$SU(3)\otimes SU(2)\otimes U(1)$.
{\bf 3)}
$\us$: There is life in the universe which is sufficiently similar to us in its biochemical requirements that
its survival criteria are approximately the same as ours.

\noindent{{\bf{Cosmological constant states}}}:\\
Our life-form can exist with a smaller than the observed CC, but not much bigger.
Let $\P(\cc,\a_1\ldots \a_n)$ be the probability distribution function (PDF) of the CC
and other parameters $\a_i$, which may include both parameters of our fundamental action,
and specifications of initial conditions. We take the CC to be defined over the range $(-m_p^4, m_p^4)$, where $m_p$ is the Planck mass (although in principle $m_p$ too could be included in the $\a_i$).
We want to compare two probabilities:
{\bf 1)}
The probability that the CC lies within a narrow range of values compatible with us. Let us be wildly generous and say that this encompasses $(-100\rho_c, 100\rho_c)$, where $\rho_c\simeq 10^{-47}~{\rm GeV}^4$ is the critical energy density,
$ P(\cc)\equiv\int_{-100\rho_c}^{+100\rho_c}
    \P(\tilde{\cc},a_1,a_2\ldots a_n)d\tilde{\cc} $,
where $a_i$ are the observed values of $\a_i$.
{\bf 2)}
The probability that the CC takes a value which is incompatible with us,
$P(\no\cc)= 1-P(\cc)$.

Putting all of the above together, we can write the following statement:
\begin{widetext}
\begin{equation}
  \frac{P(\cc)}{P(\no\cc)} =
        \frac{P(\cc|\life)P(\life)+
        P(\cc|\no\life) P(\no\life)}
        {P(\no\cc|\life)P(\life)+P(\no\cc|\no\life) P(\no\life)}
  = \frac{\Bigl( P(\cc|\us)P(\us)+P(\cc|\no\us)P(\no\us)\Bigr)+
        P(\cc|\no\life) P(\no\life)}
        { \Bigl(P(\no\cc|\us)P(\us)+P(\no\cc|\no\us)P(\no\us)\Bigr)+
        P(\no\cc|\no\life) P(\no\life)} ~.~~ \nn
\end{equation}
\end{widetext}
This ratio will become large if the observed CC is probable, and will go to zero if it isn't.
The advantage of looking at the ratio instead of the probabilities themselves is that in the ratio we may tame some of the terrible normalization issues associated with defining such probabilities over unspecified, and unspecifiedly-large, ensembles.

\noindent{{\bf{Life and us:}}} \\
As the definition of the subcategory ``us" has strong effects on subsequent calculations \cite{neal}, we now consider a few ways in which this subcategory can be defined.
The use of observers or any proxy for observers (for example, reference \cite{bousso} uses free energy as an observer proxy) as a statistical measure to determine how likely it will be to observe a parameter value depends on how the observer or proxy is defined \cite{s+t}.

A relatively relaxed way to define ``us" in the context of fundamental particle physics is to require that a life form needs the same known physics that we need, namely the standard model parameters of $SU(3)\otimes SU(2)\otimes U(1)$. Thus ``us" can be defined as
$ \{\us''\}\equiv \{ \life\}\cap \{ {\vec\a}_{sm}=
            {\vec\a}_{sm}^{(*)} \} $,
where ${\vec\a}_{sm}^{(*)}$ stands for the particular values of these parameters necessary for our existence. The parameters ${\vec \a}_{sm}$ include fermion masses, coupling constants, {\it etc}. This definition of ``us" isn't restrictive enough to enable us to say what are the general survival requirements of this subcategory. For example, ${\vec\a}_{sm}$ doesn't determine the baryon asymmetry or the dark matter density.

We can be more restrictive and define us as the intersection of intelligent life with the known parameters ${\vec\a}_{sm}$ as well as all the unknown parameters which are relevant for fundamental physics ${\vec\a}_{other}$,
$\{\us'\}\equiv \{ \life\} \cap \{ {\vec\a}_{sm}=
            {\vec\a}_{sm}^{(*)}\}\cap\{{\vec\a}_{other}=
            {\vec\a}_{other}^{(*)}\}$.
We don't know what parameters ${\vec\a}_{other}$ are, these may be the parameters of supersymmetry, string theory, {\it etc}. They may include information about initial conditions. Even though we have tightened the definition, it is still hard to make generic statements about this subcategory of life forms: for example, we can imagine compensation mechanisms between the unknown and known parameters, creating two different regions in the parameter-space in which life can be sustained. We don't know the degeneracy structure in the subspace of ${\vec\a}_{sm}$ created by existence of the unknown parameters  ${\vec\a}_{other}$.

The definition we adopt is that life is considered us-like only if we are capable of making predictions about the sufficient conditions for its existence.
This is more restrictive a definition from the ones considered above,
$ \us\subset\us'\subset\us''\subset\life $.
Indeed, this definition is a very tight one: even a minuscule change in the biological, chemical, or nuclear construct of an organism can cause dramatic changes to the conditions that were sufficient for its evolution or survival, thus shifting the life-category to which it belongs. Nonetheless, it is only this most restrictive form of ``$\us$" that can be used operationally, and for which we can say something about $P(\cc|\us)$.
For all other definitions of ``us" the strongest statement we can  justify is that ``$\cc$" and ``us" aren't mutually exclusive, based on the observations of the CC value and of our existence.

\noindent{{\bf Priors:} }\\
Imagine a universe with physics described by $n$ parameters $\a_i$, which in a particular realization take the values of $a_i$ respectively.
For brevity, we introduce the following notation:
\beq
P(a_1\ldots a_n)\equiv\int_{a_1-\Delta_1}^{a_1+\Delta_1} d\a_1\ldots \int_{a_n-\Delta_n}^{a_n+\Delta_n} d\a_{n}\P(\a_1\ldots\a_n) \nn
\eeq
where $\pm\Delta_i$ is a small range around the observed value $a_i$ of $\a_i$.
Now consider the following probabilities:
{\bf (a)} $ P(a_1\ldots a_n)$, {\bf (b)} $ P(a_1\ldots a_n|A_1\ldots A_k)$ and
{\bf (c)} $ P(a_1\ldots a_n|A_1\ldots A_p)$, where $k<p$. $A_1,A_2\ldots A_p$ are the values of a ``complete" set of observables, ${\cal A}_1, \ldots, {\cal A}_p$, {\it i.e.} if we measured all the ${\cal A}_j$ we would know the values of all the $\alpha_i$. For simplicity we assume that the observables aren't degenerate with each other, $p=n$, and that we can have a ``perfect measurement" of them,
meaning no error bars (the generalizations are straight forward).

In {\bf (a)} we ask what are the probabilities of all the parameters with flat priors. There isn't anything else we can say without additional information about the underlying theory. The answer depends, of course, on the parametrization.
We will assume that the volume of the region containing, say, the $95\%$ confidence levels for all the parameters is finite, $V_1$.
In {\bf (b)} we have perfectly measured a subset of observables, $A_1 - A_k$ which provide us with $k$ constraints, reducing the number of unconstrained parameters from $n$ to $n-k$.
If we condition the probability of the parameters with this observational information, and if the observables $A_i$ are sensitive to combinations of the parameters $\a_i$, we should expect the probability to be localized around the values $a_i$, reducing the $95\%$ confidence level volume to $V_2$, where $V_2\leq V_1$.
In {\bf (c)}, we ask what are the probabilities that the parameters take the values $a_i$, given a measurement of {\it all} observables.
In this case we can give an exact answer, $ P(a_1\ldots a_n|A_1\ldots A_n) = 1$.
So as we gather more information and apply stronger priors, the PDFs shrink into delta functions around the values which are consistent with the observables.
However, one should be careful about what information this really contains -- it tells us what parameter values are most likely in {\it this particular} realization, but it doesn't give any information about the original PDF of the theory. Tracing backwards from existing observations alone isn't a particularly effective way of constraining the underlying probability distribution of possible fundamental theories.

Coming back to the CC issue, the original probability ratio $P(\cc)/P(\no\cc)$ of the theory is unknown, though probably incredibly small because the parameter space of $\no\cc$ is vast. One can limit the question to $P(\cc|\life)/P(\no\cc|\life)$, or even further to $P(\cc|\us)/P(\no\cc|\us)$.
One may attempt to obviate this source of uncertainty by defining anthropic arguments in terms of entropy as in \cite{bousso}, but the question still comes down to the same one: namely must life exploit entropy generation under conditions comparable to what we observe today, or can it exploit it under vastly different conditions?  As shown above, it isn't surprising that the probabilities appear to grow more in favor of what we actually observe.
To take the above to extremity:
\beq
 \frac{P(\cc|\textrm{this paper})}{P(\no\cc|\textrm{this paper})}=\frac{1-\e}{\e}
  \label{proof}
\eeq
where $\e\ll 1$.
It would be wrong to take the above statement as a ``mechanism" to explain the observed value of the CC. What \eqref{proof} really tells us is that the value of $\sim 0.72$ is the most probable existing value in our realization, or to be more accurate, it reflects what we think we know. We haven't learned anything about the PDF in an ensemble of all possible universes, and certainly haven't explained its value in the causal sense of the word.

\noindent{{\bf Sampling:}} \\
We now attempt to quantify the difference between putting a condition of ``life", and putting the stronger condition of a life form which is ``us"-like.
One can imagine a parameter space containing all the parameters relevant to the emergence of life. Each point in this parameter space doesn't necessarily specify a universe; there might be additional parameters which don't affect the emergence of life but are necessary for a full description of a  universe. Over this parameter space we can define a function {\bf Life}, the probability of the emergence of life for these parameter values, given all possible universes. Our existence tells us that (modulo concerns about defining probabilities in an infinite space) {\bf Life} is non-zero at the point in the parameter space where our universe lies, but we don't know {\bf Life}'s value (no pun intended). Also, as we are the only life form and this is the only universe we know, we have no information whatsoever about how {\bf Life} extends over the rest of the parameter space. It might spread over a wide range of parameters, implying life can take many forms, or it might be a delta-function located around parameter values which lead to our life form. If the latter is true, the use of ``us" or ``life" as a condition for the CC is the same in that it statistically equates  the states ``life" and ``us". However, our limited data set which consists of a single sample of a single point doesn't probe much of the parameter space.
The fact that we know only ``us" doesn't mean that we are the only possible life form %\footnote{Notice the use of the word ``possible", not ``existing". While the question whether extra-terrestrial life exists is a fascinating one, it bears no relevance to this work.}
.

An historical  illustration of sampling bias may be called the ``New Guinea problem":
Imagine a scientist native of a small isolated New Guinea mountain village circa 1000 years ago. Looking around, she would  see that outside the hospitable valley of her village there are steep mountains covered in thick jungle. With no evidence for the existence of any  people outside her own village, our scientist would be led to identify the states of ``life" and ``from-this-village". Obviously this incorrect conclusion is due purely to sampling bias.
Today we may be in the same situation. We don't know of other life forms, which often leads to the implicit assumption that we are generic observers.
We may, for example, assume that all life exploits free energy in a way that tracks matter entropy production (which is then argued to be dominated by stellar processes \cite{bousso}), but we emphasize that we have no idea if this need be the case.
For example, it has been argued in \cite{Kephart:2002bf} that most of the entropy of the universe is in the gravitational radiation produced in the formation of the supermassive black holes at the centers of galaxies. The conclusions derived based on different sampling biases will be different \cite{ks,s+t}. While the possibility that we are typical observers exists, we don't know how to estimate how probable it is, and we aren't going to assume anything about our typicality.

Considering the possibility that life may take many forms, the conditioned probability ratio for the CC is
\begin{widetext}
\begin{equation}
  R \equiv
        \frac{P(\cc|\life)}{P(\no\cc|\life)} =
        \frac{P(\cc|\us)P(\us|\life)+P(\cc|\no\us)P(\no\us|\life)}
        {P(\no\cc|\us)P(\us|\life)+P(\no\cc|\no\us)P(\no\us|\life)}
  = \frac{\b\a+\g(1-\a)}{(1-\b)\a+(1-\g)(1-\a)} ~, \label{limits}
\end{equation}
\end{widetext}
where we have defined $\a=P(\us|\life)$, $\b=P(\cc|\us)$ and $\g= P(\cc|\no\us)$.
We now examine a few limits of Eq.~\eqref{limits}.

The case of $\a= 1$ amounts to assuming we are the only possible life form, which is sometimes an implicit assumption in anthropic arguments. We get that there is no difference between the priors of ``life" and ``us", $ R |_{\a=1}= \b/(1-\b) = P(\cc|\us)/P(\no\cc|\us) $, and $R$ peaks as $\b\rightarrow 1$. Standard Anthropic reasoning is to attempt to evaluate $\b$, and argue that $\b \simeq 1$.

The opposite limit, $\a= 0$, describes the case where we are an anomaly, and in fact most life forms are very different from us.
We then get that $R |_{\a=0}= \g/(1-\g) \neq P(\cc|\us)/P(\no\cc|\us) $.
It is interesting to notice that in this limit, $R$ peaks as $\g\rightarrow 1$.
A value of $\g$ close to $1$ means that {\it all} life forms, even if very different from us, need a CC within the same range as we do. If this is true, we can say that even though our life form isn't generic, the CC range required for our existence {\it is} generic. However, this result is independent of $\beta$ and doesn't imply that what is generic is also required. We don't have a way to estimate what is a reasonable value for $\gamma$.

\begin{figure}
  \begin{center}
    \epsfig{file=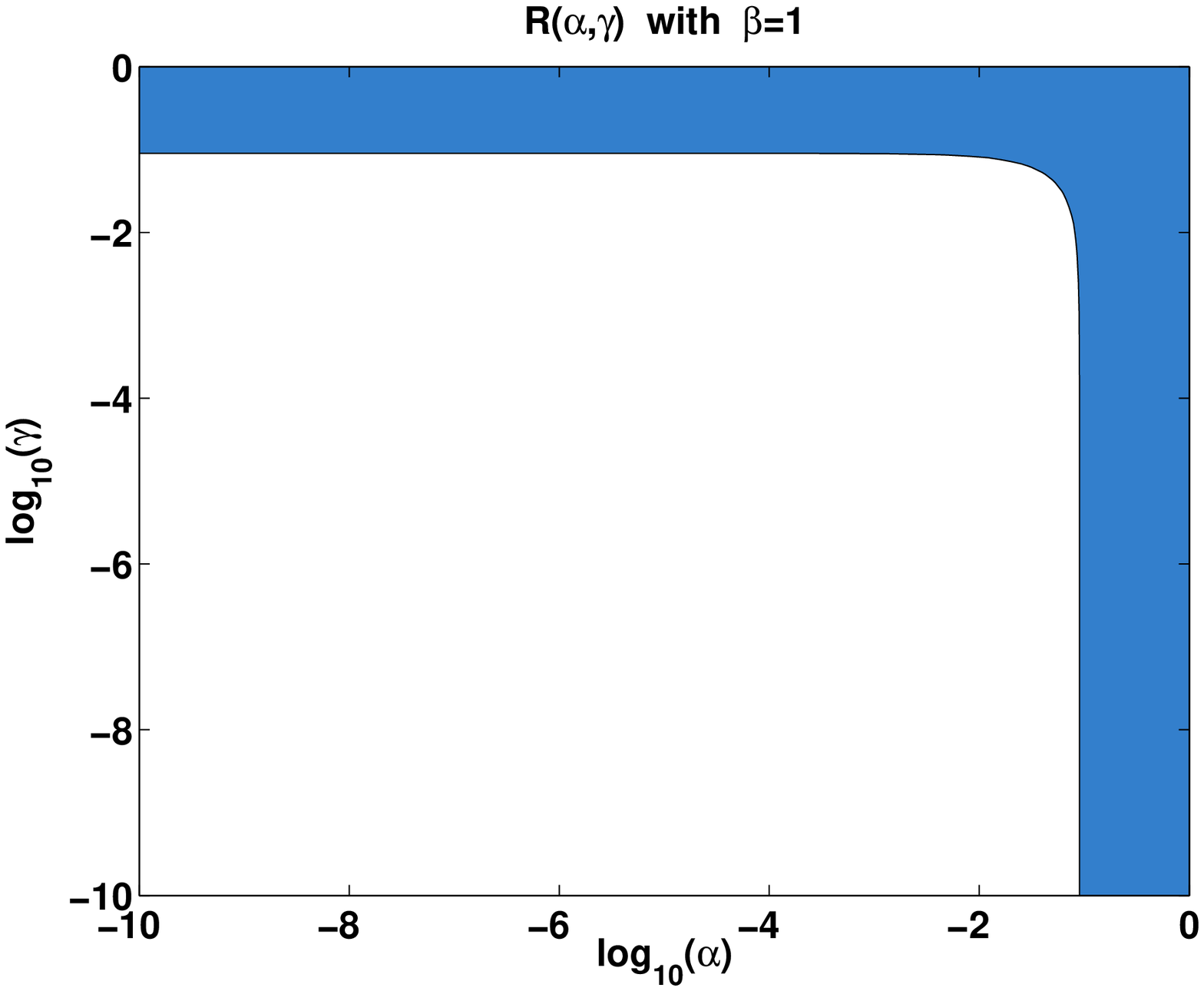,height=33mm}
    \epsfig{file=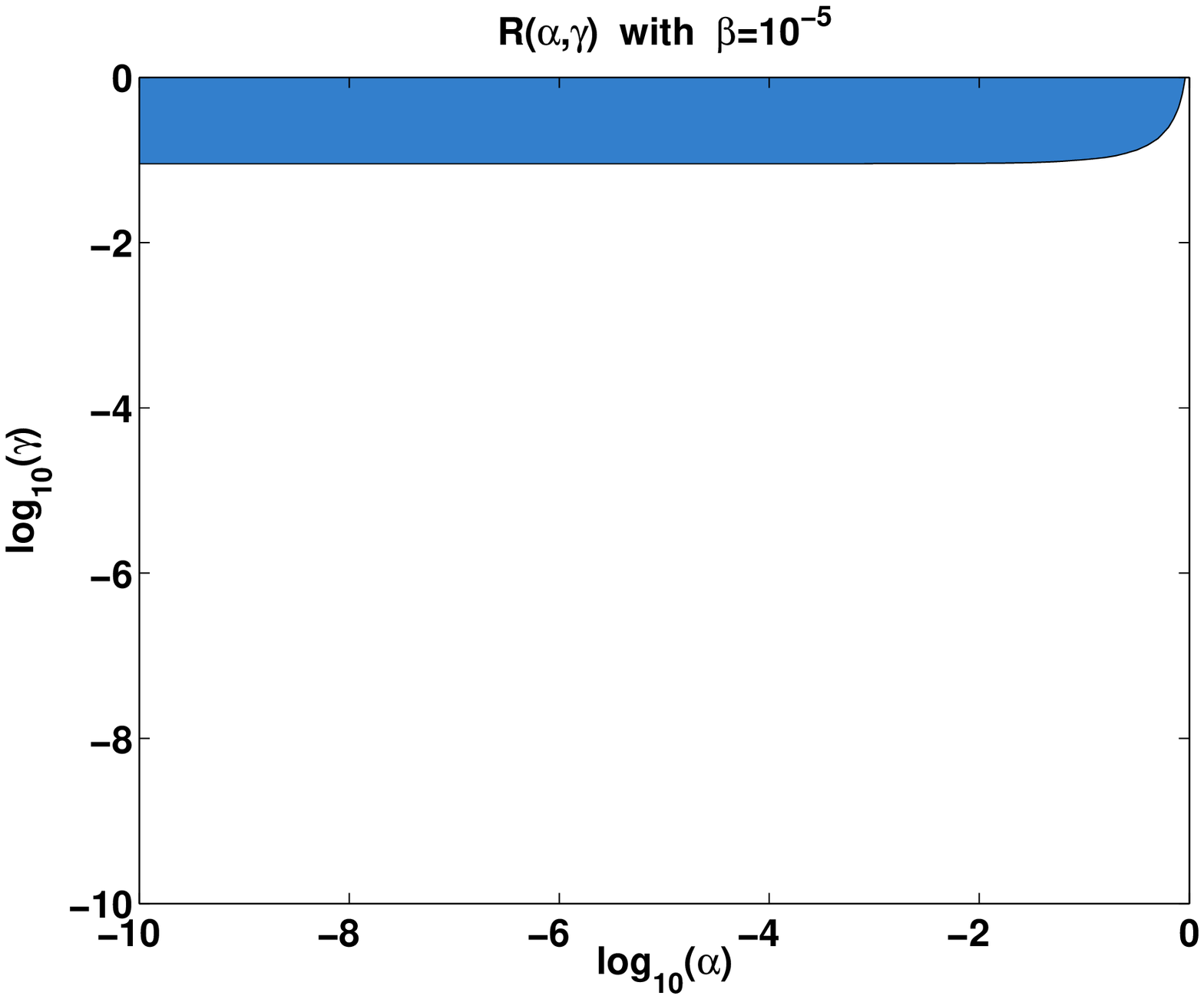,height=33mm}
  \end{center}
%  \vspace{-0.6cm}
  \caption{A contour plot of $R$ over the most ``favorable" corner
    of the $(\a,\g)$ plane, with fixed $\b=1$ (left) and
    $\b=10^{-5}$ (right).
    The shaded area indicates where $R\ge 0.1$.
  \label{lb}}
\end{figure}

The one parameter we claim to have some information about is $\beta$.
Our generous category of the allowed range for the CC together with our tight definition of ``us" are equivalent to fixing $\b=1$.
In this limit $ R |_{\b=1} = (\a+\g(1-\a))/((1-\g)(1-\a)) $. The left side of Fig.~\ref{lb} shades the area where $R\geq 0.1$ in the $(\a,\g)$ parameter space, with $\b=1$.
As we are truly ignorant about $\a$ and $\g$, we need to consider their logarithmic behavior to avoid a bias toward any order of magnitude \cite{log}. $R$ becomes substantial only as $\a\rightarrow 1$ or $\g\rightarrow 1$, and that most of the parameter space exhibits very low probability for the observed value of the CC. The right side of Fig.~\ref{lb} has $\b=10^{-5}$, which is the value that a certain weighing scheme yields \cite{s+t}.

To summarize, we can say that our existence is a good indicator of the observed CC.
That is true by construction. The probability of the observed CC given the existence of life is high only if we are the most probable life form, or if we somehow know that other life forms are as good indicators as we are of the observed CC.
%%%%%%%%%%%%%%%%%%%%%%%%%%%%%%%%%%%%%%%%%%%%%%%%%%%%%%%%%%%%%%%%%%%%%
%%%%%%%%%%%%%%%%%%%%%%%%%%%%%%%%%%%%%%%%%%%%%%%%%%%%%%%%%%%%%%%%%%%%

\noindent{{\bf Discussion:}} \\
An issue that has often caused confusion in the absence of well-defined theories is the difference between correlation and causation. The solar neutrino rate in the Davis experiment, for example, may have been correlated with the US stock market, but few physicists would argue that one had a causative effect on the other. In most cases, as we describe in this paper, anthropic reasoning can point to correlations between otherwise apparently unrelated physical parameters. However it cannot in general demonstrate causation. In this regard, there are several concepts that are often confused. Because some physical parameter is measured to take on some generic value in many cases is not equivalent to the statement that this value is {\it required}. For example, on earth, the existence of water is generically associated with life. However, there are life forms that can exist in the absence of water (see for example \cite{NAS}).
Another concept which is elusive is the difference between the attempt to explain why the CC must be small vs the attempt to explain its observed smallness. As anthropic reasoning relies on selection effects, its best hope is to give information about the probability of the {\it observed} smallness.
As we have explicitly demonstrated the statistical significance of the fact that we observe a small CC depends on how typical observers we are.
This crucial question, central to much of the current literature associated with the Landscape, remains however almost completely open, in that we currently have no scientific basis for answering it. As a result any statistical inferences are at the present time primarily wishful thinking.

Finally, the correlations illuminated by our limited anthropic understanding imply that what we ultimately learn from these arguments is that the existence of $\us$ and the existence of the observed value of $\cc$ do not contradict each other.
That is nice, but hardly surprising.

\begin{acknowledgments}
We thank R.~Trotta for comments and suggestions, and R.~Bousso and A. Linde for lively discussions.
This work was supported by the DOE and NASA at CWRU.
\end{acknowledgments}

\end{document}